\documentclass[aps,twocolumn,superscriptaddress,notitlepage]{revtex4-2}
\usepackage{graphicx}
\usepackage[T1]{fontenc}
\usepackage{xfrac}
\usepackage{upgreek}
\usepackage[usenames,dvipsnames]{xcolor}
\usepackage[linktoc=page,colorlinks,urlcolor=OliveGreen,citecolor=OliveGreen,linkcolor=OliveGreen]{hyperref}
\usepackage{amssymb,amsmath}
\usepackage{graphicx}
\usepackage{float}
\usepackage{natbib}
\usepackage{mathrsfs}
\usepackage{overpic}
\usepackage{wrapfig}
\usepackage{braket}
\usepackage{color}
\usepackage{verbatim}
\usepackage{mathtools}
\usepackage[letterpaper,textwidth=7in,top=.75in,bottom=.75in]{geometry}
\definecolor{darkblue}{rgb}{0.0 0.0 0.78}
\definecolor{darkred}{rgb}{0.5 0.0 0.0}

\newcommand{\UMDphy}{Department of Physics, University of Maryland, College Park, MD 20742, USA}
\newcommand{\QTC}{Quantum Technology Center, University of Maryland, College Park, MD 20742, USA}
\newcommand{\UMDEECS}{Department of Electrical and Computer Engineering,
University of Maryland, College Park, MD 20742, USA}

\newcommand{\Columbia}{Departamento de Física, Universidad Nacional de Colombia, sede Bogotá, Carrera 45, Colombia}

\begin{document}

\preprint{APS/123-QED}

% \title{Pulsed All-Optical Protocol for Robust Magnetometry with Nitrogen-Vacancy Center Ensembles}
\title{Noise suppression via pulsed all-optical magnetometry with nitrogen-vacancy ensembles}

\author{Xiechen Zheng}
% \thanks{These authors contributed equally to this work.}
\affiliation{\UMDEECS}
\affiliation{\QTC}
\author{Jeyson Támara-Isaza}
% \thanks{These authors contributed equally to this work.}
\affiliation{\QTC}
\affiliation{\Columbia}
% \author{Zechuan Yin}
% \affiliation{\UMDEECS}
% \affiliation{\QTC}
\author{John W. Blanchard}
\affiliation{\QTC}
% \author{Connor A. Hart}
% % \affiliation{\UMDEECS}
% \affiliation{\QTC}
% \author{\\Johannes Cremer}
% % \affiliation{\UMDEECS}
% \affiliation{\QTC}
% \author{Michael Crescimanno}
% \affiliation{\QTC}
% \affiliation{\YSUphy}
% \author{Paul V. Petruzzi}
% \affiliation{\LPS}
% \author{Matthew J. Turner}
% % \affiliation{\UMDEECS}
% \affiliation{\QTC}
\author{Ronald L. Walsworth}
\affiliation{\UMDEECS}
\affiliation{\QTC}
\affiliation{\UMDphy}

\date{\today}% It is always \today, today,

\begin{abstract}

All-optical (AO) microwave-free magnetometry using nitrogen-vacancy (NV) centers in diamond simplifies experimental design and broadens sample compatibility. 
While continuous-wave (cw) detection of AO photoluminescence (PL) changes is commonly employed, its performance is susceptible to systematic fluctuations such as optical intensity noise.
To address these challenges, we introduce a pulsed AO protocol that employs two PL measurements within an optical pulse to suppress common-mode noise.
At near-zero magnetic field, we experimentally demonstrate that the pulsed AO protocol resolves AO-PL contrast features arising from NV-NV cross-relaxation, achieving up to 10$\times$ improvement in the low-frequency noise floor compared to conventional cw AO techniques.
We further investigate the dependence of AO-PL contrast on PL readout timing and the dark time duration $\tau$ between optical pulses, with the optimal $\tau$ varying based on NV concentrations. 
These findings provide insights into optimizing NV-diamond samples for effective AO operation across diverse applications.

\end{abstract}

\maketitle
\section{Introduction}
Nitrogen-vacancy (NV) centers in diamond have emerged as a versatile platform for highly sensitive magnetometry under ambient conditions. 
Their robust optical excitation and readout capabilities, coupled with long spin coherence times at room temperature, have enabled NV centers to support a broad range of applications across the physical and life sciences \cite{taylor_high-sensitivity_2008, barry_optical_2016, glenn_high-resolution_2018, ku_imaging_2020, turner_magnetic_2020}.
These attributes also make NV centers particularly attractive for magnetic-field sensing in challenging environments \cite{sturner_integrated_2021, atallah_rapid_2022, zhu_multimodal_2025}. 

All-optical (AO), microwave-free NV magnetometry has gained increasing attention as an alternative to conventional microwave-driven schemes. 
The removal of microwave irradiation (typically used to coherently manipulate the NV spin states) offers advantages of reduced experimental complexity and potentially broader sample compatibility. 
AO magnetometry exploits magnetic-field-dependent photoluminescence (PL) changes arising from spin-state dynamics, such as spin-state mixing near 10\,mT \cite{lai_influence_2009, rondin_nanoscale_2012, tetienne_magnetic-field-dependent_2012}, cross-relaxation between NV electronic spins and substitutional nitrogen (P1) centers near 50\,mT \cite{zhang_battery_2021, lazda_cross-relaxation_2021}, and level anti-crossing within the NV electronic triplet ground states at 102.4\,mT \cite{wickenbrock_microwave-free_2016, ivady_photoluminescence_2021-1}.
Recent studies have also demonstrated AO-PL contrast features at low magnetic fields (<2\,mT), correlated with near-degenerate NV electronic spin \cite{anishchik_low-field_2015, filimonenko_weak_2020, pellet-mary_relaxation_2023, dhungel_near_2024, sengottuvel_microwave-free_2025} and hyperfine transitions \cite{zheng_all-optical_2025} from different NV orientations within the diamond host.  

Microwave-based protocols employ normalized measurements at on/off-resonance microwave frequencies to effectively suppress low-frequency common-mode noise \cite{segawa_nanoscale_2023, rubinas_interaction_2024}.
In contrast, AO measurements using continuous-wave (cw) operations are susceptible to slow systematic fluctuations, such as optical intensity noise and mechanical instability \cite{dhungel_near_2024, sengottuvel_microwave-free_2025}.
To enhance the signal-to-noise ratio (SNR) in cw AO measurements, lock-in detection is commonly employed to resolve AO-PL contrast through amplitude-modulated magnetic fields \cite{anishchik_low-field_2015, filimonenko_weak_2020, pellet-mary_relaxation_2023, zheng_all-optical_2025}. 
However, such modulated magnetic fields introduce additional design complexity, and may perturb sensitive samples such as superconducting materials \cite{kasatkin_flux_1998, sokolovsky_penetration_2004, dhungel_all-optical_2026}.

In this work, we demonstrate a pulsed AO sequence to experimentally resolve AO-PL contrast features in NV ensembles at near-zero magnetic field.
This sequence, referred to as the single-$\tau$ sequence, incorporates a fixed dark time duration $\tau$ between optical pulses; and has previously been employed with single NV spin relaxometry to improve the data acquisition speed in nanoscale magnetic noise imaging \cite{tetienne_spin_2013, pelliccione_two-dimensional_2014, schmid-lorch_relaxometry_2015, finco_imaging_2021}.
By implementing two PL readouts – one near the beginning and the other near the end of an optical pulse – we perform normalized AO measurements that suppress common-mode noise more effectively than conventional cw AO techniques.
This approach allows us to better characterize the NV-ensemble based AO-PL contrast features with common-mode noise suppression without resorting to amplitude-modulated magnetic fields required for lock-in detection. 
Furthermore, we measure the dependence of AO-PL contrast on the timing of PL readouts.
As a function of the dark time duration $\tau$ between optical pulses, we find the optimal $\tau$ for maximum of AO-PL contrast is NV-concentration dependent.

\section{Experimental Methods}
\begin{figure}[!htb]
    \centering
    \includegraphics{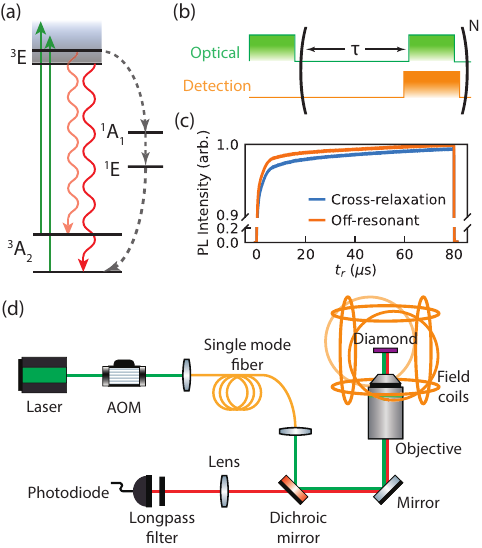}
    \caption{\label{fig:schematics}
    (a) NV energy levels and couplings allow optical initialization of electronic spin states and emission of spin-state-dependent photoluminescence (PL).
    (b) Relaxometry pulse sequence for measuring the NV electronic spin lifetime $T_1$. 
    (c) Example time-resolved PL intensity during optical illumination for resonant NV-NV cross-relaxation at near-zero magnetic field (blue) and for off-resonant conditions with an applied magnetic field (orange), respectively. The intensity of PL signals during NV-NV cross-relaxation is reduced.
    (d) Schematic of the experimental setup for near-zero-field pulsed AO magnetometry.
}
\end{figure}

The negatively-charged NV is a $C_{3v}$-symmetric color center in diamond with electronic spin triplet ($S$ = 1) ground and excited states.
Under 532\,nm optical irradiation, the NV center predominantly undergoes spin-conserving optical transitions, emitting broadband PL in the range of approximately 637–800\,nm.
Additionally, an intersystem crossing to the singlet manifold enables NVs in the excited $\ket{m_s = \pm 1}$ spin states to preferentially decay to the ground $\ket{m_s = 0}$ state with reduced PL emission. 
This process polarizes NV electronic spins to the ground $\ket{m_s = 0}$ state, creating measurable spin-state-dependent PL contrast [Fig.~\ref{fig:schematics}(a)].

\noindent

At near-zero magnetic field, NV centers aligned along the four crystallographic orientations exhibit degenerate electronic spin transition frequencies due to insignificant Zeeman shifts.
Consequently, resonant NV-NV dipolar interactions, known as NV-NV cross-relaxation, induce spin-state mixing and shorten the NV longitudinal spin relaxation time $T_1$ \cite{pellet-mary_relaxation_2023}.
The time-resolved PL intensity from an NV ensemble can be measured over time $t_r$ following optical illumination, as shown in Fig.~\ref{fig:schematics}(c).
NV-NV cross-relaxation at near-zero magnetic field reduces the measured PL intensity compared to the off-resonant case, where the spin transition degeneracy is lifted by an applied magnetic field.
By tuning the magnetic field, AO-PL contrast features associated with NV-NV cross-relaxation and the NV hyperfine interaction can be resolved \cite{zheng_all-optical_2025}.

$T_1$-based relaxometry is commonly employed for NV magnetic noise sensing \cite{tetienne_magnetic-field-dependent_2012}.
The typical relaxometry pulse sequence, shown in Fig.~\ref{fig:schematics}(b), consists of an initial optical pulse to initialize the NV ground electronic spin states to $\ket{m_s = 0}$, which subsequently relax towards a mixture of $\ket{m_s = 0, \pm 1}$ after a variable dark time $\tau$. 
A second optical pulse is applied for PL read out to determine the resulting electronic spin state; and also to optically polarize the NV for the next of $N$ repetitions of the sequence. 

In this study, we use an acoustic-optic modulator (AOM) to switch 532-nm laser light for both cw and pulsed AO-PL measurements; the light is then passed through a single-mode fiber and an objective with $\text{NA} = 0.45$, creating a Gaussian beam profile of radius $\approx30\,\mu$m at the NV ensemble in the diamond sample. 
A high sampling rate data acquisition device records the time-resolved AO-PL from a high bandwidth photodiode.
Three-axis Helmholtz coils provide controllable magnetic fields aligned along the diamond's [110], [$\bar{1}$10], and [001] crystallographic axes [Fig.~\ref{fig:schematics}(d)]. 
Table~\ref{tab:samples} summarizes the characteristics of the three CVD-grown NV-diamond samples employed in this work. 
Further details of the experimental setup are given in Appendix~\ref{app-sec:exp-details}.

\begin{table}[H]
\centering
\caption{\label{tab:samples} 
Diamond samples used in this study. All samples are electronic grade plates (few mm on each side and about 0.5\,mm thick) enriched with $^{14}$N, with a 10\,$\mu$m-thick surface layer of enhanced nitrogen and NV concentration ([N] = 16\,ppm, > 99.995\% $^{12}$C) as reported by Element Six Ltd. 
}
\begin{ruledtabular}
% \resizebox{\textwidth}{!}{%
\begin{tabular}{cc}
\textbf{Sample \#}  & \textbf{{[}NV{]} (ppm)}\\
D1                       & $\approx 3.8$ \\
D2                       & $\approx 2$   \\
D3                       & $\approx 0.3$ \\
\end{tabular}%
% }
\end{ruledtabular}
\end{table}

\section{Results}
\label{sec:results}
\begin{figure}[!htb]
\centerline{\includegraphics{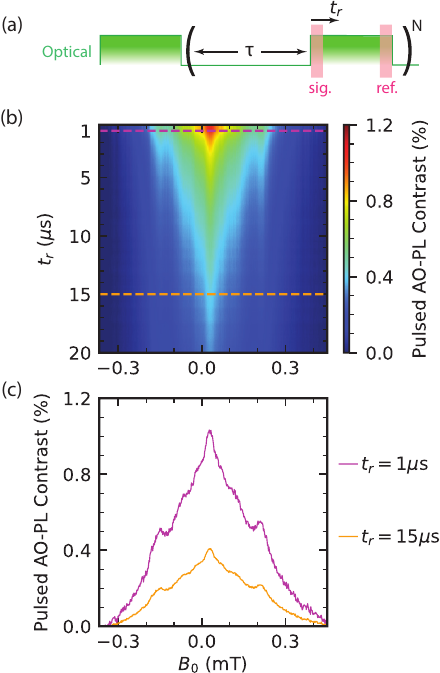}}
\caption{\label{fig:readout-timing}
(a) Protocol for normalized measurements of pulsed AO-PL intensity with fixed dark time $\tau$. Signal readout window is at variable time $t_r$; and reference readout is at the end of the optical pulse.
(b) Experimentally determined pulsed AO-PL contrast for sample D1 as a function of signal readout timing $t_r$ and applied magnetic field $B_0$ along [$\bar{1}$10]. 
(c) Line cuts of pulsed AO-PL contrast data for $t_r=1\,\mu$s and $t_r=15\,\mu$s, corresponding to horizontal dashed lines in (b), purple and orange, respectively.
Longer $t_r$ yields weaker normalized AO-PL contrast.
}
\end{figure}

We perform normalized measurements of pulsed AO-PL intensity, collected in signal (\(I_{\text{sig}}\)) and reference (\(I_{\text{ref}}\)) readout windows, each with a length of 1\,$\mu$s, within an optical pulse [Fig.~\ref{fig:readout-timing}(a)]. 
\(I_{\text{sig}}\) is measured at $t_r$ from the beginning of the optical pulse and \(I_{\text{ref}}\) is measured during the final 1\,$\mu$s of the pulse.
We choose the optical pulse length to be $80\,\mu$s, which is sufficient to achieve steady-state NV spin polarization under the applied 275\,mW laser power used in this work.
This AO-PL normalization allows for suppression of common-mode noise that equally affects both measurements, such as relatively slow variations in optical intensity.
The readout window length of 1\,$\mu$s is selected to apply a moving average filter for transient PL measurements, integrating neighboring time-resolved PL intensities to improve SNR while balancing AO-PL contrast. 
The normalized AO-PL intensity (\(n = \overline{I_{\text{sig}}} / \overline{I_{\text{ref}}}\)) at a given (near-zero) magnetic field is further compared to the value at a reference off-resonant magnetic field $B_0=0.45$\,mT (\(n^\prime = \overline{I_{\text{sig}}^\prime} / \overline{I_{\text{ref}}^\prime}\)) where the NV-NV cross-relaxation effect is minimal.
The pulsed AO-PL contrast is then calculated as \(C = 1-n/n^\prime\).

Using this pulsed measurement protocol, we experimentally determine the pulsed AO-PL contrast for sample D1 as a function of $B_0$ along [$\bar{1}$10] and $t_r$, with fixed dark time $\tau = 2\,$ms [Fig.~\ref{fig:readout-timing}(b)]. 
The results with $t_r$ between 0 and 0.5\,$\mu$s are omitted due to low PL intensities shortly after optical excitation, leading to noisier resolved AO-PL features [see Appendix~\ref{app-sec:early-readout}]. 
Fig.~\ref{fig:readout-timing}(c) shows examples of the measured near-zero-field AO-PL contrast as a function of $B_0$, at fixed $t_r = 1\,\mu$s and $t_r = 15\,\mu$s. 
The observed AO-PL contrast peaks in Fig.~\ref{fig:readout-timing}(c) result from the NV hyperfine interaction with the $^{14}$N nuclear spin, which provides additional electronic spin depolarization channels and splits the NV-NV cross-relaxation resonance near zero fields \cite{zheng_all-optical_2025}.
In addition, the measured pulsed AO-PL contrast decreases as $t_r$ increases, for all values of $B_0$. 
As a consequence, higher pulsed AO-PL contrast is achieved with the \(I_{\text{sig}}\) readout occurring shortly after optical excitation and \(I_{\text{ref}}\) readout towards steady-state.
This result is consistent with the measured transient NV-PL in Fig.~\ref{fig:schematics}(c), where larger reductions in PL intensity are seen during NV-NV cross-relaxation for times closer to the activation of optical excitation rather than at steady-state. 

\begin{figure}[!htb]
    \centering
    \includegraphics{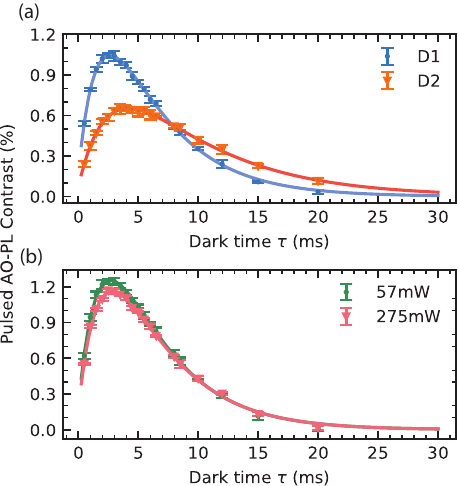}
    \caption{\label{fig:contrast_darktime}
    (a) Pulsed AO-PL contrast at $B_0=0.03\,$mT, 275\,mW laser power, and $t_r=1\,\mu$s, as a function of dark time $\tau$ for samples D1 ([NV] $\approx$ 3.8\,ppm) and D2 ([NV] $\approx$ 2\,ppm). 
    Markers are the experimental contrast values $C(\tau)$, with measurement uncertainty indicated.
    Solid lines are from a bi-exponential fit to experimental $C(\tau)$ values. 
    For both samples, the fit yields two characteristic timescales $T_{cr}$ and $T_{in}$ that can be associated with NV-NV cross-relaxation and intrinsic depolarization, respectively.
    (b) Pulsed AO-PL contrast at $B_0=0.03\,$mT and $t_r=1\,\mu$s as a function of dark time $\tau$ for sample D1 ([NV] $\approx$ 3.8\,ppm) at two laser powers, 57\,mW and 275\,mW. The two fit timescales are consistent within error bars.
}
\end{figure}

We next measure the pulsed AO-PL contrast as a function of dark time $\tau$ at fixed $t_r=1\,\mu$s and $B_0=0.03\,$mT, i.e., near the maximum contrast from the near-zero-field AO-PL feature.
The pulsed AO-PL contrast $C(\tau)$, from both sample D1 ([NV] $\approx$ 3.8\,ppm) and sample D2 ([NV] $\approx$ 2\,ppm), initially increase with $\tau$, reach a maximum value, and then decrease at longer $\tau$ [Fig.~\ref{fig:contrast_darktime}(a)]. 
For $\tau$ ranging from 0.5 to 20\,ms, we fit these experimental results quantitatively using a bi-exponential function:
\begin{equation} \label{eqn:bi-exp-fit}
    C(\tau) = A_\text{cr}e^{-t/T_\text{cr}} + A_\text{in}e^{-t/T_\text{in}}.
\end{equation} 
Here, we attribute $T_\text{cr}$ and $T_\text{in}$ to the characteristic timescales of NV-NV cross-relaxation and intrinsic depolarization, respectively, with $A_\text{cr}$ and $A_\text{in}$ being the corresponding amplitudes. 
For sample D1, $A^\text{D1}_\text{cr} = -2.69 \pm 0.38\,$, $A^\text{D1}_\text{in} = 2.86 \pm 0.41\,$, $T^\text{D1}_\text{cr} = 1.77 \pm 0.18\,$ms and $T^\text{D1}_\text{in} = 4.81 \pm 0.34\,$ms; and for sample D2, $A^\text{D2}_\text{cr} = -1.76 \pm 0.4\,$, $A^\text{D2}_\text{in} = 1.83 \pm 0.42\,$, $T^\text{D2}_\text{cr} = 2.75 \pm 0.39\,$ms and $T^\text{D2}_\text{in} = 7.36 \pm 0.85\,$ms.
These results indicate that stronger dipolar interactions from higher [NV] shortens both timescales, creating higher maximum pulsed AO-PL contrast with a shorter dark time $\tau$.
In addition, the experimentally determined AO-PL contrast change as a function of $\tau$ is consistent with the measured differences of NV spin $T_1$ between on-resonance ($B_0 = 0.03\,$mT) and off-resonance ($B_0 = 0.45\,$mT) magnetic fields (see Appendix~\ref{app-sec:T1-diff}).

We next repeat the measurements on sample D1 for different laser powers, 57\,mW and 275\,mW [Fig.~\ref{fig:contrast_darktime}(b)].
The optical pulse is extended from the previous value of 80\,$\mu$s to 300\,$\mu$s to ensure sufficient optical excitation at the end of the pulse for both laser powers.
Fitting the data to the bi-exponential function in Eq.~\ref{eqn:bi-exp-fit} yields $A_\text{cr} = -3.56 \pm 0.91$ ($-3.64 \pm 1.19$), $A_\text{in} = 3.75 \pm 0.96$ ($3.8 \pm 1.23$), $T_\text{cr} = 1.88 \pm 0.3\,$ms ($2.04 \pm 0.36\,$ms) and $T_\text{in} = 4.66 \pm 0.52\,$ms ($4.65 \pm 0.6\,$ms) at 57\,mW (275\,mW), respectively.
For the two laser powers, the fitted $T_\text{cr}$ and $T_\text{in}$ values are in agreement within error bars, consistent with these timescales being primarily dependent on sample properties such as [NV].
The modest observed reduction in maximum AO-PL contrast ($\sim$0.1\%) at 275\,mW may result from NV ionization and charge cycling, which decrease the population of negatively-charged NVs by converting them to the neutral charge state (NV$^0$) \cite{edmonds_characterisation_2021, giri_selective_2019, cardoso_barbosa_impact_2023}.

\begin{figure}[!htb]
    \centering
    \includegraphics{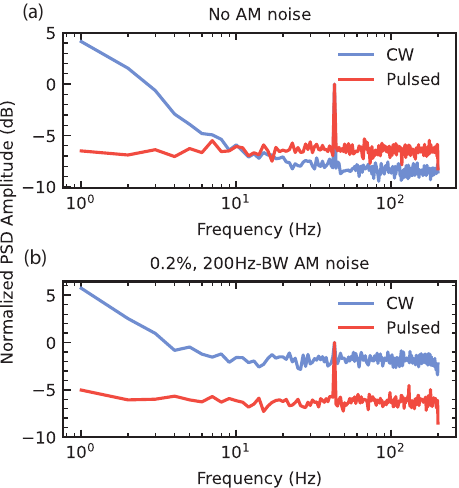}
    \caption{\label{fig:noise_supp}
    Normalized power-spectral-density (PSD) relative to a test signal (1\,$\mu$T, 43\,Hz) for cw and pulsed AO measurements of sample D1.
    (a) No AM noise added to the AOM RF power.
    (b) Additional AM noise with 0.2\% depth and 200\,Hz bandwidth.
}
\end{figure}
Finally, we compare the noise suppression performance of pulsed AO to conventional cw AO operation for sample D1 without the use of lock-in detection.
We start by measuring 30 1-s time traces at $B_0 = 0.05\,$mT where the AO magnetometer is magnetically sensitive.
For pulsed AO operation, we set the optical pulse duration to $80\,\mu$s for shorter time spacing between \(I_{\text{sig}}\) and \(I_{\text{ref}}\) readout timings, and keep the total sequence duration at $2.5\,$ms (i.e., sampling rate of 400\,S/s and Nyquist frequency of 200 Hz); for cw AO operation, we use a sampling rate of 100\,kS/s. 
Each measured 1-s time trace is converted into a power-spectral-density (PSD) plot using the calibrated slope of the AO-PL contrast feature at $B_0 = 0.05\,$mT, before being averaged. 
A fixed test signal with an amplitude of $1\,\mu$T at $f_\text{test} = 43\,$Hz is applied for both pulsed and cw AO operation.
The averaged PSDs are subsequently normalized with respect to the PSD amplitude at $f_\text{test}$.
This normalization allows us to compare the noise suppression performance in units of decibels (dB), with 0\,dB at $f_\text{test}$ due to the PSD normalization. 
To ensure both pulsed and cw AO operation are optimally sensitive to the applied test signal, we apply laser power of 275\,mW for pulsed and 4\,mW for cw AO measurements. 
Modest cw optical power is needed for AO-PL measurements utilizing NV-NV cross-relaxation, in order to balance the optical pumping and spin relaxation processes \cite{zheng_all-optical_2025}.

We achieve $\sim10\,$dB improvement in the near DC frequency range from pulsed relative to cw AO operation, with diminishing noise suppression capability $>10\,$Hz, as shown in Fig.~\ref{fig:noise_supp}(a).
Above 10\,Hz, the cw AO method exhibits a lower noise floor, primarily due to its much higher sampling rate (100\,kS/s).
In contrast, pulsed AO operation, with an effective sampling rate of 400\,S/s, is susceptible to aliasing, whereby high-frequency noise components fold back into the 0–200\,Hz band, elevating the noise floor in the 10–200\,Hz region \cite{heinzel_spectrum_2002, kirchner_aliasing_2005}.

We introduce additional optical intensity noise via amplitude modulation (AM) to the AOM RF power, with a 0.2\% depth and 200\,Hz bandwidth from the noise modulation function.
It is worth noting that the AM depth is defined relative to the AOM RF power, which varies between cw and pulsed AO operation modes to achieve different average laser powers.
This approach allows us to characterize the noise suppression performance under realistic conditions, including the effects of relative optical intensity noise.
The 0.2\% AM modulation depth is chosen in order to obtain a minimally measurable test signal for cw AO operation.
A comparison of measured near-zero-field AO-PL contrast features with AM between the two operation modes is given in Appendix~\ref{app-sec:noise-feature-comparison}.
In Fig.~\ref{fig:noise_supp}(b), the pulsed AO normalization method shows a similar $\sim10\,$dB improvement near DC similar compared to the case without added optical intensity noise [Fig.~\ref{fig:noise_supp}(a)].
Additionally, pulsed AO normalization maintains a consistently 4-5\,dB lower noise floor across the higher-frequency (10-200\,Hz) range relative to the cw AO method.
These results demonstrate the robustness of the proposed pulsed AO magnetometry protocol in effectively rejecting common-mode noise compared to the conventional cw AO technique. 

\section{Outlook}
As demonstrated above, pulsed AO magnetometry using an NV ensemble near zero magnetic fields provides a significant ($\sim10\times$) improvement in low-frequency noise floor relative to conventional AO techniques.
This advantage exploits normalized measurements from two AO-PL readouts within an optical pulse (the single-$\tau$ protocol); and is particularly promising for sensitive AO DC magnetometry in the absence of amplitude-modulated magnetic fields.
While this work focused on the NV-NV cross-relaxation effect for AO magnetometry, alternative AO-PL contrast mechanisms and operational regimes, such as spin-state mixing (see Appendix~\ref{app-sec:spin-mixing}), can be also implemented using this protocol. 

The measured variation in AO-PL contrast as a function of sequence dark time $\tau$ reflects the interplay between NV-NV cross-relaxation and intrinsic depolarization processes.
These dynamics collectively determine the optimal $\tau$, which varies with NV concentration for the maximum AO-PL contrast.
In this study, the NV-diamond sample with higher [NV] ($\approx 3.8$\,ppm) exhibits stronger AO-PL contrast at a shorter $\tau$ of about 2.5\,ms.

However, this millisecond-scale $\tau$ results in a low duty cycle for the pulsed AO sequence used here, emphasizing the need for further optimization to achieve higher sampling rates and enable sensitive pulsed AO measurements of higher-frequency magnetic signals.
Employing NV-diamond samples with high [N] presents a potential solution.
In bulk samples, [N] can reach $\sim100\,$ppm, enabling [NV] $\sim50\,$ppm and reducing $T_1$ to $\sim70\,\mu$s \cite{choi_depolarization_2017}.
While such higher [N] and [NV] may improve AO-PL contrast and pulsed AO sequence duty cycle by shortening both $T_\text{cr}$ and $T_\text{in}$ timescales, this approach also introduce trade-offs that require further evaluation for pulsed AO optimization, including: broader AO-PL linewidths and reduced effectiveness of common-mode noise rejection from longer optical spin polarization times due to higher optical absorption.

Our findings provide new insights on NV-diamond sample selection and AO magnetometry protocols relevant to low size, weight, and power (low-SWaP) quantum sensors.
Such devices can be particularly well suited for sensitive NV magnetometry in challenging environments, including extreme temperature, pressure, corrosive, and radiation conditions, broadening the diversity of quantum sensing applications \cite{fu_sensitive_2020}.

\section*{Acknowledgments}
We thank Connor A. Hart for technical insights on implementing the pulsed AO protocol and helpful feedback on the manuscript; and Johannes Cremer and Michael Crescimanno for valuable discussions. 
We acknowledge the U.S. Army Research Laboratory for providing NV diamond samples.
This work is supported by, or in part by, the Department of Energy under Grant. No. DESC0021654; U.S. Army Research Laboratory under Contract No. W911NF2420143; and the University of Maryland Quantum Technology Center.

\appendix
\renewcommand{\theequation}{S\arabic{equation}}
\renewcommand\thefigure{S\arabic{figure}}    
\setcounter{figure}{0}  
\section{Experimental details} 
\label{app-sec:exp-details}
532\,nm cw laser light (Lighthouse Photonics Sprout-H-10W) is modulated with an AOM (G\&H 3250-220), after which the first order diffracted beam is transmitted through a single-mode optical fiber (Thorlabs P3-460B-FC-5).
The modulated laser beam is then focused through a microscope objective (Olympus 20X/NA 0.45) to create a Gaussian excitation laser beam profile of radius $\approx$ 30\,$\mu m$ to optically excite NV centers in the diamond samples under study. 
NV-diamond sample properties are summarized in Table~\ref{tab:samples} of the main text. 
NV photoluminescence (PL) is collected through the same objective, directed with a dichroic mirror (Semrock FF552-DI02), filtered with a long-pass filter (Semrock BLP01-532R-25), and focused onto a high-bandwidth avalanche photodiode (Thorlabs APD430A) through a 25.4\,mm lens (Thorlabs LA1951-B-ML). 

A three-channel power supply (Rohde \& Schwarz HMP4030) connected to relay boards (Winford RLY202-5V-DIN) is used to drive a three-axis Helmholtz coil array (Ferronato BH300-3-A) to provide a bias magnetic field of controllable magnitude and direction.
Calibration of applied current to bias magnetic fields is in the Appendix of Ref.~\cite{zheng_all-optical_2025}.
The NV-diamond sample is mounted on a piezoelectric rotational stage (Thorlabs ELL14) to independently align the applied magnetic field with respect to the NV crystallographic axes, with $B_0 \parallel [\bar{1}10]$, and $B_1 \parallel [100]$ and $B_2 \parallel [010]$ aligned separately.

The output of the photodiode used to collect NV PL is sent to a high-speed data-acquisition device (Gage RazorMax CSE50216) for time-resolved measurements with a 200\,MS/s sampling rate.
Measurement pulse sequences are synthesized by a pulse generator (Swabian Instruments Pulse Streamer 8/2): channel 0 provides TTL pulses to a solid state switch (Mini-Circuits ZASW-2-50DRA+) to gate the AOM; and channel 1 synchronizes the data acquisition on Gage RazorMax.
A signal generator (Stanford Research Systems SG384) provides the RF signal at 225\,MHz that, together with an RF amplifier (Mini-Circuits ZHL-03-5WF+), drives the AOM. 
The RF amplitude from the signal generator to the AOM is varied to control the laser power delivered to the NV-diamond sample.
The same signal generator is used to provide amplitude-modulated RF generation via the noise modulation function to introduce additional optical intensity noise to the experiment.

\section{AO-PL features from short signal readout timing ($t_r$)}
\label{app-sec:early-readout}
\begin{figure}[!htb]
    \centering
    \includegraphics{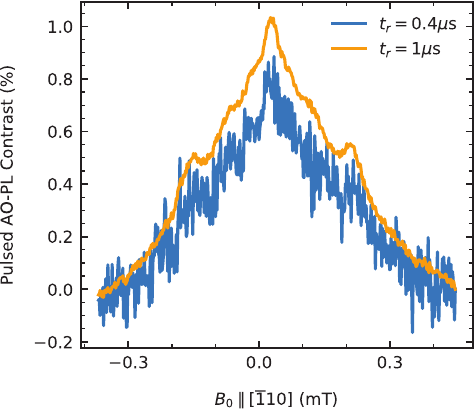}
    \caption{\label{sup-fig:early_readout}
    Pulsed AO-PL contrast of sample D1 measured for signal readout pulse values $t_r=0.4\,\mu$s and $t_r=1\,\mu$s, with dark time $\tau = 2\,$ms. $t_r < 1\,\mu$s yields weaker AO-PL contrast and worse SNR due to poor optical spin polarization after the optical pulse.
}
\end{figure}
Fig.~\ref{sup-fig:early_readout} shows a comparison of measured near-zero-field pulsed AO-PL contrast features for sample D1, with fixed dark time $\tau = 2\,$ms, and for signal readout pulse values $t_r = 0.4\,\mu$s and $t_r = 1\,\mu$s.
The result with $t_r = 0.4\,\mu$s exhibits weaker AO-PL contrast and worse SNR.
The weaker AO-PL contrast is attributed to poor optical spin polarization shortly after the optical pulse, as a longer $\tau$ allows the NV electronic spins to relax toward thermal equilibrium.
In addition, the lower PL emission intensity lead to worse SNR.
For $t_r$ between $0\,\mu$s and $0.4\,\mu$s, the near-zero-field AO-PL contrast features are no longer identifiable in measurement with our setup.

\section{Differences of polarization lifetime ($T_1$) measurements between on- and off-resonance magnetic fields}
\label{app-sec:T1-diff}
\begin{figure}[!htb]
    \centering
    \includegraphics{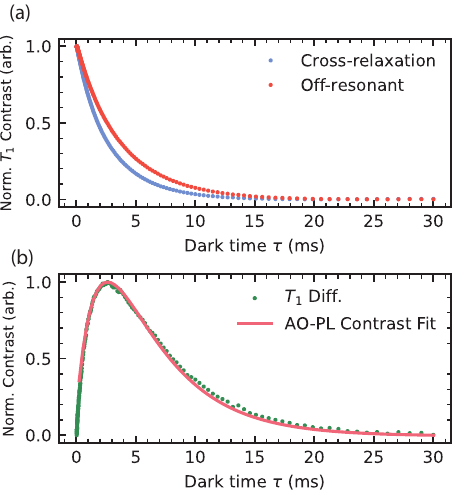}
    \caption{\label{sup-fig:T1_diff}
    (a) Experimentally-determined NV ground-state spin-polarization lifetime ($T_1$) of sample D1 as a function of dark time $\tau$. 
    Fits of pulsed AO-PL decay (not shown) yield $T_1 = 2.602\pm0.008\,$ms with cross-relaxation magnetic field at $B_0 = 0.03\,$mT, and $T_1 = 3.70\pm0.012\,$ms with off-resonant magnetic field at $B_0 = 0.45\,$mT. 
    (b) Differences of $T_1$ results in Fig.~\ref{sup-fig:T1_diff}(a) and a bi-exponential fit to the measured AO-PL contrast of sample D1 in Fig.~\ref{fig:contrast_darktime}(a) as a function of dark time $\tau$, with good consistency found between the methods.
}
\end{figure}
Fig.~\ref{sup-fig:T1_diff}(a) presents the experimentally-determined NV ground-state spin-polarization lifetime ($T_1$) of sample D1 as a function of dark time $\tau$ from 0 to 30\,$\mu$s, using the relaxometry pulse sequence shown in Fig.~\ref{fig:schematics}(b).
The normalized pulsed AO-PL $S(\tau)$ measurements are fit with a stretched exponential function (not shown in Fig.~\ref{sup-fig:T1_diff}(a)) with a stretched factor $\beta$ as:
\begin{equation}
  S(\tau) = A(1 + \exp(-(\frac{\tau}{T_1})^\beta))
\end{equation}
When NV-NV cross-relaxation is strongest at magnetic field $B_0 = 0.03\,$mT, the fit yields $T_1 = 2.602 \pm 0.008\,$ms, with $\beta = 0.926 \pm 0.004$.
At the off-resonant magnetic field $B_0 = 0.45\,$mT, $T_1 = 3.7 \pm 0.012\,$ms, with $\beta = 0.976 \pm 0.004$.

Fig.~\ref{sup-fig:T1_diff}(b) shows the normalized difference between these two $T_1$ relaxometry measurements alongside the normalized bi-exponential fit of pulsed AO-PL contrast as a function of $\tau$ presented in Fig.~\ref{fig:contrast_darktime}(a, blue). 
The good agreement between these normalized results indicate that the optimal $\tau$ for achieving maximum AO-PL contrast is linked to the differences of $T_1$ between cross-relaxation and off-resonant conditions, and is therefore dependent on the NV concentration.

\section{Comparison of near-zero-field AO-PL contrast features between pulsed and cw operations with additional optical intensity noise}
\label{app-sec:noise-feature-comparison}
\begin{figure}[!htb]
    \centering
    \includegraphics{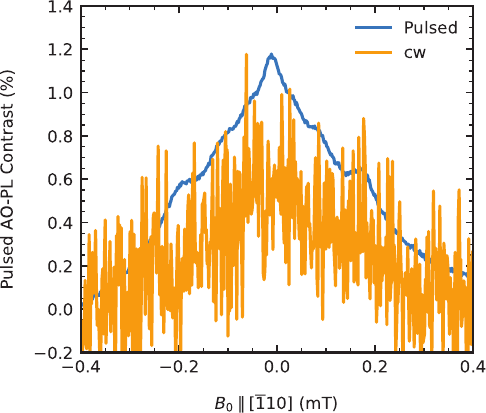}
    \caption{\label{sup-fig:pulsed_vs_cw}
    Measured near-zero-field AO-PL contrast of sample D1 using pulsed (blue) and cw (orange) methods as a function of $B_0$. Additional optical intensity noise is applied through AM with 1\% depth and 200\,Hz bandwidth from noise modulation of the AOM RF power. The pulsed method can still resolve clear AO-PL contrast features under substantial optical intensity noise.
}
\end{figure}

Fig.~\ref{fig:noise_supp}(b) in the main text demonstrates the enhanced noise suppression achieved with pulsed AO operation, using a test signal at $f_\text{test} = 43\,$Hz with additional amplitude modulation (AM) noise at 0.2\% depth and 200\,Hz bandwidth applied to the AOM RF power.
To further evaluate the noise suppression performance, we increase the AM depth to 1\% and compare the measured near-zero-field AO-PL contrast features of sample D1 for cw and pulsed operation, as shown in Fig.~\ref{sup-fig:pulsed_vs_cw}.
The wall time for both measurements is kept approximately identical to ensure a fair comparison, as cw AO measurements have a higher sampling rate and average more data points at each magnetic field. 
The results show that while the cw technique fails to resolve AO-PL contrast features with good SNR under strong optical noise, the pulsed technique maintains good robustness.
This result highlights the advantage of the pulsed method for resolving AO-PL contrast features during applied magnetic field sweeps, which are typically slower than measuring an oscillating (AC) magnetic signal at a fixed bias field.

\section{Pulsed AO-PL contrast from spin-state mixing}
\label{app-sec:spin-mixing}

\begin{figure*}[!htb]
    \centering
    \includegraphics{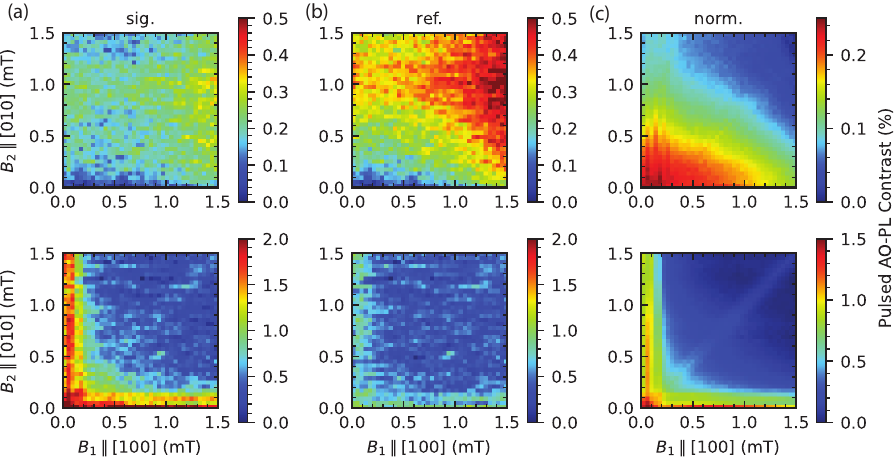}
    \caption{\label{sup-fig:spinMixing}
    Pulsed AO-PL contrast as a function of applied magnetic field magnitude and direction, for: (a) signal readout pulse value $t_r = 1\,\mu$s; (b) reference readout pulse value $t_r = 79\,\mu$s; and (c) after normalization, using sample D3 with $\approx \,$0.3\,ppm [NV] (top) and sample D1 with $\approx \,$3.8\,ppm [NV] (bottom).
    Implementing two PL readout normalization with sample D3 allows the characterization of pulsed AO-PL contrast from both spin-state mixing as a non-negligible background and weak NV-NV cross-relaxation as stripelike features.
    Unlike pulsed AO-PL contrast from NV-NV cross-relaxation, which is primarily observed in the signal readout measurement (a, bottom), a significant portion of pulsed AO-PL contrast from spin-state mixing appears in the reference readout measurement (b, top).
    Future studies will explore optimizing AO pulse sequences for specific operational conditions.
}
\end{figure*}
\noindent

Fig.~\ref{sup-fig:spinMixing} shows experimental results for pulsed AO-PL contrast as a function of $B_1 \parallel [100]$ and $B_2 \parallel [010]$, using sample D3 with approximately 0.3\,ppm NV concentration (top) and sample D1 with approximately 3.8\,ppm NV concentration (bottom), respectively.
We vary both $B_1$ and $B_2$ from 0 to 1.5\,mT, with step size approximately 0.04\,mT. 
The lower NV concentration in sample D3 allows for the measurement of AO-PL contrast from NV electronic spin states mixing under applied magnetic fields that are not aligned with the NV quantization axis, as the AO-PL contrast arising from NV-NV cross-relaxation is much weaker \cite{zheng_all-optical_2025}.
We plot the pulsed AO-PL contrast from \(I_{\text{sig}}\) at $t_r = 1\,\mu$s [Fig.~\ref{sup-fig:spinMixing}(a)], \(I_{\text{ref}}\) at $t_r = 79\,\mu$s [Fig.~\ref{sup-fig:spinMixing}(b)], and normalized values using the steps described in Sec.~\ref{sec:results} [Fig.~\ref{sup-fig:spinMixing}(c)].
Each two-dimensional AO-PL contrast plot is normalized to the maximum intensity from a given scan of ($B_1$, $B_2$).
By implementing two PL readout normalization with sample D3, we successfully measure pulsed AO-PL contrast from spin-state mixing as a non-negligible background, along with contributions from weak NV-NV cross-relaxation as stripelike features [Fig.~\ref{sup-fig:spinMixing}(c, top)].
A stronger background spin-state mixing pulsed AO-PL contrast is observed in the reference readout measurement [Fig.~\ref{sup-fig:spinMixing}(b, top)] compared to the signal readout measurement [Fig.~\ref{sup-fig:spinMixing}(a, top)].
This result differs from that of sample D1, where pulsed AO-PL contrast arising from NV-NV cross-relaxation is primarily observed in the signal readout window [Fig.~\ref{sup-fig:spinMixing}(a, bottom)], as described in Sec.~\ref{sec:results} of the main text.
These findings suggest that the pulsed AO technique can be adapted to measure AO-PL contrast using various measurement protocols under a wide range of applied magnetic fields.
Optimizing AO pulse sequences for specific operational conditions will be explored in future studies.

\end{document}